\documentclass[aps,pra,twocolumn,showpacs,amsmath,amssymb,amsfonts,floatfix,]{revtex4}

\usepackage[pdftex]{graphicx}
\usepackage{color}
\usepackage{textcomp}
\usepackage{epsfig}
\usepackage{bm}
\usepackage{graphics}
\usepackage{comment}
\usepackage{ulem}

\begin{document}

\title{Low-Temperature Linear Thermal Rectifiers Based on Coriolis forces}
\author{Suwun Suwunnarat$^1$, Huanan Li$^{1}$, Ragnar Fleischmann$^{2}$ and Tsampikos Kottos$^{1,2}$}
\affiliation{\mbox{$^1$Department of Physics, Wesleyan University, Middletown, Connecticut 06459, USA}\\
\mbox{$^2$Max Planck Institute for Dynamics and Self-organization (MPIDS),  37077 G\"{o}ttingen, Germany}}
\begin{abstract}
We demonstrate that a three-terminal harmonic symmetric chain in the presence of a Coriolis force, produced by a rotating
platform which is used to place the chain, can produce thermal rectification. The direction of heat flow is reconfigurable and
controlled by the angular velocity $\Omega$ of the rotating platform. A simple three terminal triangular lattice is used to 
demonstrate the proposed principle.
\end{abstract}
\
\maketitle

\section{Introduction}

In the last few years considerable research effort has been invested in developing appropriately engineered structures that display 
novel transport properties not found in nature. In the thermal transport framework, this activity has recently started to gain a lot of 
attention. Apart from the purely academic reasons, there is a growing consensus on its practical implications in the efforts of the 
society to manage its energy resources efficiently. Thus research programs that aim to propose new methods, designs or materials 
that allow to harness and mold the heat flow at the nanoscale level in ways that can affect society's energy consumption needs 
are at the forefront of the research agenda. Some of the targets that are within our current nanotechnology capabilities include the 
generation of nanoscale heat-voltage converters, thermal transistors and rectifiers, nanoscale radiation detectors, heat pumps and 
even thermal logic gates \cite{LLP03,D08,COGMZ08,NGPB09,LRWZHL12,ZL10}. 
 
In spite of these efforts the understanding of thermal transport and more importantly the manipulation of heat current is still in its 
infancy. This becomes more obvious if one compares with the tremendous achievements of the last 
fifty years in understanding and managing electron transport. In contrast, for example, the important problem of thermal rectification
is addressed only by a handful of researchers. The majority of these proposals relies on the interplay of system non-linearities with 
structural asymmetries \cite{WL08,LWC06,WL07}. At the nanoscale limit where the phonon 
mean-free path is comparable to the size of the devices the inherent nonlinearities are irrelevant and transport is fully dominated 
by ballistic phonon transport (ballistic limit). However recently it was shown in Ref. \cite{ZWL10,P10,MWDL10} that in the ballistic
regime one can achieve thermal rectification under very specific conditions: low temperature limit associated with quantum harmonic 
systems in the nonlinear response domain, and the presence of a third reservoir which acts as a probe and it is coupled asymmetrically with 
the rest of the system in order to break any spatial symmetry. The importance of asymmetry of the system itself was further highlighted
in Ref. \cite{P10}.

In this paper we focus on the development of a new concept for the creation of unidirectional thermal valves that control and direct 
heat currents on the nanoscale level. Specifically we propose to create a ballistic symmetric thermal rectifier which relies on Coriolis forces. 
In contrast to previous studies our proposal does not rely on any structural asymmetry. Rather, an asymmetry is induced in a very controllable 
way by the Coriolis force allowing a high flexibility in the rectification properties without a change of the structural setup. The system 
consists of a symmetric harmonic lattice placed on a rotating platform with angular velocity $\Omega$. The direction of heat current is 
controlled on the sign of $\Omega$, thus allowing for a 
reconfigurable heat current management. Using the non-equilibrium Green's function formalism we derive the conditions of optimal
operation of the structure. Subsequently we have verified these predictions via detailed numerical calculations using a simple variant
of the proposed structure. 

The paper is organized as follows. In section \ref{sec:model} we present the theoretical model. The mathematical formalism
for the evaluation of heat current and the conditions for non-reciprocity are discussed in section \ref{sec:form}. Finally a numerical
confirmation of our predictions are given in section \ref{sec:numerics}. Our conclusions are given in the last section \ref{sec:conclusions}.

\section{Theoretical Model }\label{sec:model}

We consider a harmonic lattice of $N$ particles coupled together with spring constants $k^C$. All particles are assumed to have the same 
mass $m$. Below, without any loss of generality, we assume that all masses are $m=1$. The lattice is placed on a platform in the $X-Y$ 
plane. Thus each particle can move on this platform and has $D=2$ 
degrees of freedom. The connectivity of the lattice is uniquely defined by the ${\cal N}=N*D$ dimensional symmetric force matrix $K^C$. We 
will assume the most general case where the coupling is not necessarily confined to be only between nearest (in space) masses. Moreover we 
will assume that all masses are coupled harmonically to a central post which is parallel to the $Z$ axis and  placed in the middle of the platform.

The lattice rotates around the post with a constant angular velocity $\overrightarrow {\Omega}$. It is useful to describe the motion of the
system in the rotating frame. In this frame the particles are characterized by their equilibrium position vector $R^0=(R_{1}^0,\cdots,R_{N}^0)^T$, 
by their displacements vector $S=(S_{1}^{x},S_{1}^{y},S_{1}^{z},\cdots,S_{N}^{x},S_{N}^{y},S_{N}^{z})^{T}$ and the associated 
conjugate canonical momenta vector $p_{C}=(p_{C1}^{x},p_{C1}^{y},p_{C1}^{z},\cdots,p_{CN}^{x},p_{CN}^{y},p_{CN}^{z})^{T}$ \footnote{The 
superscript $T$ indicates transposition}. The Hamiltonian that describes the system in the rotating frame takes the form:
\begin{equation}
H_{C}=\frac{1}{2}p_{C}^{T}p_{C}+\frac{1}{2}S^{T}K^{C}S-\left(R^{0}+S\right)^{T}A\, p_{C}\label{eq:hamilt}
\end{equation}
where $A$ is the ${\cal N}\times{\cal N}$ block-diagonal matrix
\begin{equation}
A={\rm diag}\{{\tilde{A}}_{D}\};{\tilde{A}}_{D=3}=\begin{bmatrix}{\tilde{A}}_{D=2} & 0\\
0 & 0
\end{bmatrix};{\tilde{A}}_{D=2}=\begin{bmatrix}0 & \Omega\\
-\Omega & 0
\end{bmatrix}\label{eq:Amatrix}
\end{equation}
with the property $A^{T}=-A$. The last term in Eq. \eqref{eq:hamilt} describes the Coriolis force in the rotating frame.  

The rotating lattice described by Eq.\eqref{eq:hamilt} is coupled with three equivalent co-rotating heat baths. The latter are always attached 
to the same three particles of the lattice $\alpha=R_1,R_2,R_3$. We will assume that two of these baths are at high $T_{\rm hot}$ and low 
$T_{\rm cold}$ constant temperatures, respectively. The third bath acts as a probe and its temperature $T_{R_3}=T_p$  is adjusted self-consistently  
such that the net heat flux from it is zero.  In general this can lead to different $T_p$ values in the forward ($T_{R_1}=T_{\rm hot}$ and $T_{R_2}=T_{\rm cold}$) and backward ($T_{R_1}=T_{\rm cold}$ and $T_{R_2}=T_{\rm hot}$) process.

For simplicity, the heat baths are described quasi-classically, i.e. we promote the relative momenta $p_{\alpha}$ of the $\alpha-$bath particles 
and their displacements $u_{\alpha}$, with respect to the rotating frame, to conjugate canonical pairs \cite{LK15}. Nevertheless, 
we assume that the statistical properties of the phonons in these thermal reservoirs are not affected by the Coriolis force and respect quantum 
statistics \cite{Ming2010}. The three baths, each consisting of an infinite number of harmonically coupled bath particles, are discribed by Hamiltonians
\begin{equation}
H_{\alpha}=\frac{1}{2}p_{\alpha}^{T}p_{\alpha}+\frac{1}{2}u_{\alpha}^{T}K^{\alpha}u_{\alpha},\,\alpha=R_{1},R_{2},R_{3}\label{eq:bath}
\end{equation}
with a (semi-infinite) harmonic force matrix $K^{\alpha}=\left(K^{\alpha}\right)^{T}$ contains additional $-\left|\overrightarrow{\Omega}\right|^{2}\cdot I$
terms due to the centrifugal force ($I$ denotes the semi-infinite identity matrix). 

Finally we have assumed that each particle is pinned to the substrate via a quadratic potential with the coupling constant $k_0$. This 
pinning potential guarantees that the lattice 
particles have an equilibrium position and it can be incorporated easily in all force matrices $K^{C},K^{\alpha}$, see Eqs.~\eqref{eq:hamilt} and 
\eqref{eq:bath}.

We are now ready to write down the total Hamiltonian of the bath-lattice system which takes the form:
\begin{equation}
H_{tot}=H_{C}+\sum_{\alpha}H_{\alpha}+\sum_{\alpha}H_{\alpha C}
\label{eq:total}
\end{equation}
where $H_{\alpha C}=u_{\alpha}^{T}V^{\alpha C}u_{C}$ describes the coupling between the lattice particles and the heat baths. 

Below we will be assuming that the initial condition of the total system is a direct product state 
\begin{equation}
\label{initialstate}
\hat{\rho}_{ini}\left(t_{0}\right)=\prod_{\alpha}\otimes\frac{e^{-H_{\alpha}/\left(k_{B}T_{\alpha}\right)}}{\mathrm{Tr}\left(e^{-H_{\alpha}/
\left(k_{B}T_{\alpha}\right)}\right)}\otimes\frac{e^{-H_{C}/\left(k_{B}T_{C}\right)}}{\mathrm{Tr}\left(e^{-H_{C}/\left(k_{B}T_{C}\right)}\right)}
\end{equation}
which after sufficient long time will be relaxing to a steady state $\hat{\rho}^{ss}$. Note that the temperature of the two baths $T_{\alpha}$ are 
kept constant and the steady-state thermal current will not depend on the initial temperature of the lattice model $T_{C}$. The value of the 
temperature of the probe in the NESS will be determined by the temperature of the other baths.

\section{Formalism}\label{sec:form}

\subsection{Heat Current}\label{ssec:cur}

We want to calculate the steady-state thermal current flowing from the hot reservoir $T_{\rm hot}$ towards the cold reservoir $T_{\rm cold}$.
To this end we will employ the standard nonequilibrium Green's function technique~\cite{Wang2013}. The steady-state current out of the heat bath $\alpha$ is
\begin{eqnarray}
I_{\alpha} &\equiv& -\mathrm{Tr}\left[\hat{\rho}^{ss}\frac{d\hat{H}_{\alpha}\left(t\right)}{dt}\right]\nonumber \\
 & = & \sum_{\gamma\neq\alpha}\int_{0}^{\infty}\frac{d\omega}{2\pi}\hbar\omega{\cal T}_{\gamma\alpha}\left[\omega\right]
\left(f_{\alpha}-f_{\gamma}\right).
\label{eq:Landauer-like}
\end{eqnarray}
where $f_{\alpha}=f(\omega,T_\alpha)=\left\{ \exp\left(\hbar\omega/k_{B}T_{\alpha}\right)/-1\right\} ^{-1}$ is the Bose distribution associated with the 
particles at the bath $\alpha$ which have fixed temperature $T_{\alpha}$ and ${\cal T}_{\gamma\alpha}$ is the transmission coefficient
from the $\alpha$-th bath to the $\gamma$-th bath. The hot and cold reservoirs,  with temperatures $T_{\rm hot}$ and $T_{\rm cold}$, 
will be attached to the particles $R_1,R_2$ while the third - probe - reservoir with varying temperature $T_p$ will be always attached to 
the $R_3$-particle.

During the forward process we attach the hot reservoir to the particle $R_1$ so that $T_{R_1}=T_{\rm hot}$ and the cold reservoir, to the 
particle $R_2$ so that $T_{R_2}=T_{\rm cold}$.  The associated heat current $I_f$ out of bath $R_2$ is then given from Eq. (\ref{eq:Landauer-like})
\begin{equation}
\label{for}
I_{f} 
 = \sum_{\gamma=R_1,R_3}\int_{0}^{\infty}\frac{d\omega}{2\pi}\hbar\omega{\cal T}_{\gamma,R_2}\left[\omega\right]
\left(f(T_{R_2})-f(T_{\gamma})\right) .
\end{equation}
where the bath temperature $T_{p}^f$ associated with the probe reservoir has to be evaluated from the requirement that the net 
current $I_p$ flowing in the probe reservoir is zero i.e.
\begin{equation}
\label{forT}
I_{p} 
 = \sum_{\gamma=R_1,R_2}\int_{0}^{\infty}\frac{d\omega}{2\pi}\hbar\omega{\cal T}_{\gamma,p}\left[\omega\right]
\left(f(T_{p}^f)-f(T_{\gamma})\right) =0.
\end{equation}
The backward process is associated with the reverse configuration i.e. $T_{R_2}=T_{\rm hot}$ and $T_{R_1}=T_{\rm cold}$. The associated 
heat flux $I_b$ out of bath $R_2$ and the corresponding temperature of the probe $T_p^b$ are derived using similar relations as the ones shown in Eqs. (\ref{for},\ref{forT}).

\subsection{The Transmission Coefficient}\label{ssec:trans}

We proceed with the evaluation of the transmission coefficient ${\cal T}_{\gamma\alpha}$. The latter is needed for the calculation of the
forward and backward currents, and therefore for the evaluation of the rectification parameter $R$ via Eq. (\ref{rectif}). The transmission coefficients can be expressed as
\begin{equation}
\label{transmission}
{\cal T}_{\gamma\alpha}\left[\omega\right]=\mathrm{Tr}\left[G_{CC}^{r}\Gamma_{\gamma}G_{CC}^{a}\Gamma_{\alpha}\right];\quad
\Gamma_{\alpha}\equiv i\left[\Sigma_{\alpha}^{r}-\Sigma_{\alpha}^{a}\right]
\end{equation}
where $G_{CC}^{r(a)}$ is the the retarded (advanced) Green's functions of the lattice. $\Sigma_{\alpha}^{r\left(a\right)}=
\left(V^{\alpha C}\right)^{T}g_{\alpha}^{r(a)}V^{\alpha C}$ is the associated self-energy which can be expressed in terms of the
equilibrium Green's function $g_{\alpha}^{r(a)}$ of the isolated heat bath $\alpha$ and the coupling matrix $V^{\alpha C}$. Moreover using the definition Eq.~(\ref{transmission}) we 
can deduce that $\Gamma_{\alpha}$ is a symmetric matrix i.e.\ $\Gamma_{\alpha}^{T}=\Gamma_{\alpha}$.

The retarded Green's function $G_{CC}^{r}$ for the central junction in the frequency domain takes the form
\begin{equation}
\label{g_cc_r}
G_{CC}^{r}\left[\omega\right]=\left[\left(\omega+i0^{+}\right)^{2}-K^{C}-\Sigma^{r}\left[\omega\right]-A^{2}-2i\omega A\right]^{-1}
\end{equation}
where $\Sigma^{r}=\sum_{\alpha}\Sigma_{\alpha}^{r}$ denotes the total retarded self-energy
due to the interaction with all the heat baths. The associated advanced Green function can be expressed in terms of $G_{CC}^{r}$ as
\begin{equation}
\label{g_cc_a}
G_{CC}^{a}[\omega]=\left(G_{CC}^{r}\left[\omega\right]\right)^{\dagger}.
\end{equation}
Using expression Eq. (\ref{g_cc_r}) we can deduce that 
\begin{equation}
\label{Gsym}
G_{CC}^{r\left(a\right)}\left[\omega,-\Omega\right]=G_{CC}^{r\left(a\right)}\left[\omega,\Omega\right]^{T}
\end{equation}.

Using the cyclic property of the trace, the symmetric nature of further $\Gamma_{\alpha}^{T}=\Gamma_{\alpha}$ and Eq. (\ref{Gsym}),
we can easily show that 
\begin{equation}
\label{Tsym}
{\cal T}_{\gamma\alpha}\left[\omega,-\Omega\right]={\cal T}_{\alpha\gamma}\left[\omega,\Omega\right]\neq {\cal T}_{\gamma \alpha}\left[\omega,\Omega\right]
\end{equation}
which lead us to the conclusion that for $\Omega\neq 0$ the time-reversal symmetry of the system is broken due to the rotation. 

In order to highlight the importance of the Coriolis force in thermal rectification we will be considering set-ups which involve structural 
rotational symmetry. In this case, any asymmetric heat transport phenomena that we will find are due to Coriolis force only (contrast this with the
studies of Refs. \cite{WL08,LWC06,WL07} where thermal rectification is due to structural asymmetries). As an outcome of this rotational symmetry 
assumption we have that the transmission coefficients ${\cal T}_{\alpha,\gamma}$ satisfy the following relations
\begin{equation}
\label{Trelations}
{\cal T}_{R_{2}R_{1}}={\cal T}_{R_{3}R_{2}}={\cal T}_{R_{1}R_{3}}, \; {\cal T}_{R_{1}R_{2}}={\cal T}_{R_{2}R_{3}}={\cal T}_{R_{3}R_{1}}.
\end{equation}
The above expressions can allow us to calculate the transmission coefficient ${\cal T}_{\gamma\alpha}[\omega]$, and thus to proceed 
with the evaluation of the forward and backward currents $I_f, I_b$.

\subsection{The Rectification Parameter}\label{ssec:rectif}

Below we will quantify the rectification effect by the rectification parameter $R$ defined as
\begin{equation}
\label{rectif}
R=\frac{-\Delta I}{\max\left\{ \left|I_f\right|,\left|I_{b}\right|\right\} };\quad \Delta I=I_{f}+I_{b}.
\end{equation}
When $\Delta I>0\,\left(R<0\right)$ the system acts as a thermal rectifier favoring the direction from the 
bath placed at $R_{2}$ to the bath attached at $R_{1}$ while the reverse is true in case that $\Delta I<0\,(R>0)$.
We will assume  $T_{\rm hot/\rm cold}=T_0 \pm \Delta T$ and analyze the dependence of $R$ on the temperature difference $\Delta T$.

Using Eqs. (\ref{eq:Landauer-like},\ref{forT},\ref{Trelations}) we can evaluate the rectification parameter $R$ up to the second 
order in the temperature difference $\Delta T$. The current difference $\Delta I$ is given by~\cite{Ming2010}:
\begin{equation}
\Delta I =(1-a^{2})\frac{K}{F\left({\cal T}_{R_{2}R_{3}}\right)+F\left({\cal T}_{R_{1}R_{3}}\right)}  \Delta T^{2}\label{eq:delta_I}
\end{equation}
where $a, F$ and $K$ are defined as follows: the parameter $a$ is defined as
\begin{equation}
\label{aparameter}
a  = \frac{F\left({\cal T}_{R_{1}R_{3}}\right)-F\left({\cal T}_{R_{2}R_{3}}\right)}{F\left({\cal T}_{R_{1}R_{3}}\right)+F\left({\cal T}_{R_{2}R_{3}}\right)}
\end{equation}
and its absolute value is always smaller than unity i.e. $|a|\leq 1$. Therefore the coefficient $(1-a^2)$ appearing in Eq. (\ref{eq:delta_I})
is always greater than zero i.e. $(1-a^2)>0$. The positive definite functions $F\left({\cal T}_{\alpha R_{3}}\right)>0$ are defined as
$F\left({\cal T}_{\alpha R_{3}}\right) =  \int_{0}^{\infty}\frac{d\omega}{2\pi}\hbar\omega\left(\frac{\partial f\left(\omega,T\right)}{\partial T}\right)_{T_{0}}
{\cal T}_{\alpha R_{3}},\quad \alpha=R_{1},\, R_{2}$. Finally $K$ in Eq. (\ref{eq:delta_I}) is defined as
\begin{widetext}
\begin{eqnarray}
K  &=&  \int_{0}^{\infty}\frac{d\omega_{1}}{2\pi}\int_{\omega_{1}}^{\infty}\frac{d\omega_{2}}{2\pi}W(\omega_1,\omega_2,T_0)\mathcal{T}(\omega_{1},\omega_{2});\label{kparameter}\\
W &=&\frac{\hbar^4\omega_{1}\omega_{2}}{k_BT_0^2}\left(\frac{\partial f\left(\omega_{1},T\right)}{\partial T}\frac{\partial f\left(\omega_{2},T\right)}{\partial T}\right)_{T_{0}}\left\{\omega_{2}\left(1+2f\left(\omega_{2},T_{0}\right)\right)-\omega_{1}\left(1+2f\left(\omega_{1},T_{0}\right)\right)\right\}\nonumber
\end{eqnarray}
\end{widetext}
with the weight function $W$ and $\mathcal{T}(\omega_{1},\omega_{2}) =  {\cal T}_{R_{2}R_{3}}\left[\omega_{1}\right]{\cal T}_{R_{2}R_{3}}\left[\omega_{2}\right]
\left\{ \frac{{\cal T}_{R_{1}R_{3}}\left[\omega_{2}\right]}{{\cal T}_{R_{2}R_{3}}\left[\omega_{2}\right]}-\frac{{\cal T}_{R_{1}R_{3}}\left[\omega_{1}\right]}{{\cal T}_{R_{2}R_{3}}\left[\omega_{1}\right]}\right\}$. 

From Eq. (\ref{eq:delta_I}) and the subsequent discussion we can draw a number of conclusions: Since the linear term in the $\Delta T$ expansion is
zero we immediately conclude that in the linear-response regime or in the classical limit our system has zero rectification parameter, i.e. $R=0$,and 
therefore one needs to consider the quantum regime beyond linear response. Second, we realize that since $1-a^{2}>0$ and $F\left({\cal T}_{\alpha R_{3}}
\right)>0$, the sign of $\Delta I$ (and thus the sign of the rectification parameter $R$) is completely determined by the sign of $K$, see Eq. (\ref{kparameter}).
Below we will be confirming these predictions for a simplified version of our model Eq. (\ref{eq:total}).

\section{An Example of Thermal Rectification due to Coriolis Force}\label{sec:numerics}

\begin{figure}
\includegraphics[width=1\linewidth, angle=0]{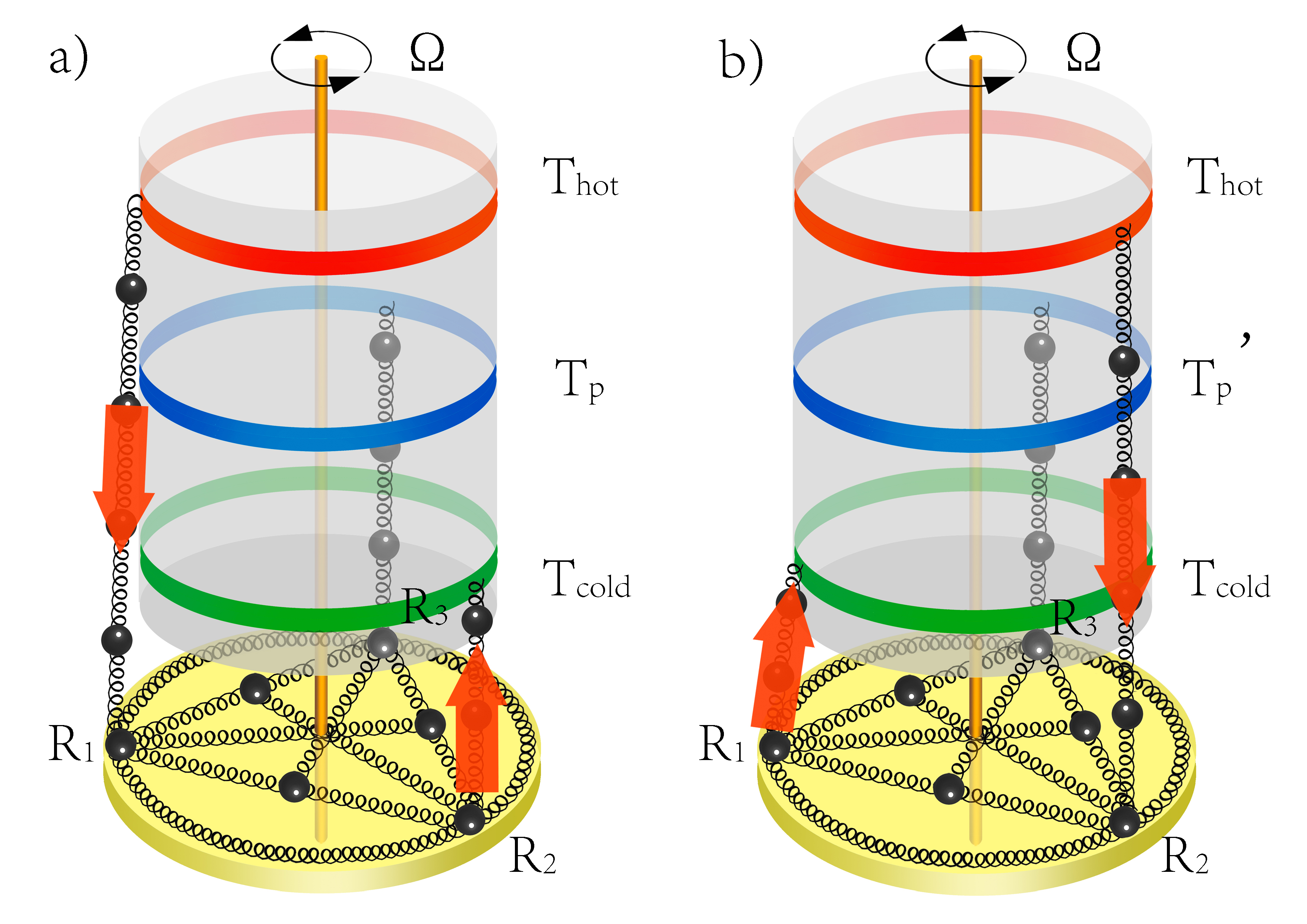}
\caption{(Color online) A schematic of the thermal rectifier due to rotation when $N_b=1$: on each edge of the triangle masses are coupled together with equal harmonic
springs up to the next-nearest neighbor. The masses are also attached to a post with similar springs and they move on a platform at the $X-Y$ plane which rotates
with a counterclockwise (CCW) angular velocity $\Omega$. a) The mass $R_1$ is coupled to a high-temperature bath with temperature $T_{R_1}=T_{hot}$ while the  
mass $R_2$ is coupled to the low-temperature bath with temperature $T_{R_2}=T_{cold}$. The mass $R_3$ is connected to the probe bath with temperature $T_{R_3}
=T_{p}$. b) The mass $R_1$ is coupled to a low-temperature bath with temperature $T_{R_1}=T_{cold}$ while the  mass $R_2$ is coupled to the high-temperature 
bath with temperature $T_{R_2}=T_{hot}$. The mass $R_3$ is again connected to the probe bath with a different temperature $T_{R_3}=T_{p}'$.}
\label{fig1}
\end{figure}

We will demonstrate the rectification effect, induced by the Coriolis force $(\Omega\neq 0)$, using a simple version of the general model, 
Eq. (\ref{eq:total}). The underlying {\it symmetric} harmonic lattice has an equilibrium configuration associated with an equilateral triangle. 
Each of the three corners of the triangle $R_{1}^{0},\, R_{2}^{0},\, R_{3}^{0}$ are occupied by equal masses $R_{1},\, R_{2},\, R_{3}$, coupled 
with the same harmonic coupling $k_{C}$ to a post which is placed at the center $O$ of the triangle. On every edge of the triangle, say for
example edge $R_{1}^{0}R_{2}^{0}$, (by excluding the masses at the vertexes) there are additional $N_{b}$ equal masses, which are connected
with the center $O$ of the triangle by springs dividing the angle $\angle R_{1}^{0}OR_{2}^{0}$ equally. An equal harmonic coupling up to the 
next-nearest neighbor is considered between the masses on each edge of the triangle. Furthermore, the masses $R_{1},\, R_{2},\, R_{3}$ are 
attached to three equivalent 1D Rubin baths~\cite{Rubin1971}. The 1D Rubin baths $\alpha=R_{1},R_{2},R_{3}$ are made up of a semi-infinite
spring chain with $K_{nm}^{\alpha}=\delta_{nm}(2k^{\alpha}-\Omega^{2}+k_{0})-k^{\alpha}\delta_{n\pm1,m}$. Here an additional coupling 
$k_{0}$ with the substrate is assumed. An illustration of our model with $N_{b}=1$ is shown in Fig.~\ref{fig1}.

The symmetric force matrix $K^{C}$ in Eq.~\eqref{eq:hamilt} can be obtained systematically as follows. Each spring (harmonic coupling) will contribute to the matrix 
$K^{C}$ with an additive harmonic term. These spring contributions can be classified in three categories. First, we consider the coupling between two masses $R_{j}$ 
and $R_{k}$ of the system. We use  local Euclidean coordinate system to describe the relative displacement of the particles with respect to their equilibrium positions. 
The local Euclidean coordinate system for each particle is specified in the following way: its origin is taken as the equilibrium position of the particle and its x-axis and 
y-axis  respectively  coincide with the radial ($r$) and tangential direction ($\theta$) of fixed equilibrium position of the specific particle.
Specifically, using this local Euclidean coordinate system,  the contribution of the spring connecting the particles $R_{j},R_{k}$ to the matrix $K^{C}$ is
\begin{eqnarray}
[K^{C}]_{jr_{j},jr_{j}} & = & [K^{C}]_{j\theta_{j},j\theta_{j}}=[K^{C}]_{kr_{k},kr_{k}}=[K^{C}]_{k\theta_{k},k\theta_{k}}=k_{C}\nonumber\\
{}[K^{C}]_{jr_{j},kr_{k}} & = & [K^{C}]_{kr_{k},jr_{j}}=-k_{C}\cos(\theta_{j}^0-\theta_{k}^0)\nonumber\\
{}[K^{C}]_{jr_{j},k\theta_{k}} & = & [K^{C}]_{k\theta_{k},jr_{j}}=-k_{C}\sin(\theta_{j}^0-\theta_{k}^0)\nonumber\\
{}[K^{C}]_{j\theta_{j},kr_{k}} & = & [K^{C}]_{kr_{k},j\theta_{j}}=k_{C}\sin(\theta_{j}^0-\theta_{k}^0)\nonumber\\
{}[K^{C}]_{j\theta_{j},k\theta_{k}} & = & [K^{C}]_{k\theta_{k},j\theta_{j}}=-k_{C}\cos(\theta_{j}^0-\theta_{k}^0).
\end{eqnarray}
where $\left(r_{j}^0,\,\theta_{j}^0\right)$ and $\left(r_{k}^0,\,\theta_{k}^0 \right)$ is the polar coordinate for the equilibrium positions  $R_{j}^0$ and $R_{k}^0$ of two masses $R_{j}$ 
and $R_{k}$ .
Second, we evaluate the contribution to the matrix $K^{C}$ due to the spring coupling between the mass $R_{j}$ and the center $O$. We have that
$[K^{C}]_{jr_{j},jr_{j}}=[K^{C}]_{j\theta_{j},j\theta_{j}}=k_{C}$. Similarly the coupling between the mass $R_{j}$ and the platform provides the contribution $[K^{C}]_{jr_{j},jr_{j}}=[K^{C}]_{j\theta_{j},j\theta_{j}}=k_{0}$. Third, due to the coupling with the bath $\alpha$ additional terms are added to the matrix  $K^{C}$. For example, for $\alpha=R_{1}$ they are $[K^{C}]_{1r_{1},1r_{1}}=[K^{C}]_{1\theta_{1},1\theta_{1}}=k^{R_{1}}$.

Finally the nonzero elements of the coupling matrices (see below Eq. (\ref{eq:total}) for a definition) are respectively
$\left[V^{R_{1}C}\right]_{1,1r_{1}}=-k^{R_{1}},\left[V^{R_{2}C}\right]_{1,2r_{2}}=-k^{R_{2}}$
and $\left[V^{R_{3}C}\right]_{1,3r_{3}}=-k^{R_{3}}$. In the following,
$k^{R_{1}}=k^{R_{2}}=k^{R_{3}}=k$ is assumed for simplicity. 
 
The thermal rectification effect is present when the (structurally) symmetric system is rotating with an angular velocity $\Omega$ around the post situated
at the center $O$. To be specific, first the bath coupled to the mass $R_{1}$ is set to the higher temperature $T_{R_{1}}=T_{hot}=T_{0}+\Delta T$ and the bath 
coupled to the mass $R_{2}$ is set to the lower temperature $T_{R_{2}}=T_{cold}=T_{0}-\Delta T$ (forward process). The temperature $T_{p}$ of the probe bath coupled to 
the mass $R_{3}$ is determined by the zero flux condition as discussed above. We then calculate the forward thermal current $I_f$ using Eq.~\eqref{for}. Second, by 
reversing the temperature bias, $i.e.,$ $T_{R_{2}}=T_{hot}$ and $T_{R_{1}}=T_{cold}$, and adjusting the temperature of the probe bath $R_{3}$ to be $T_{p}'$, 
the thermal current $I_b$ for the backward process is evaluated. In the following natural units are assumed, i.e.\ $\hbar=1,\, m=1,\, k=1,\, k_{B}=1$.

\begin{figure}
\includegraphics[width=1\linewidth, angle=0]{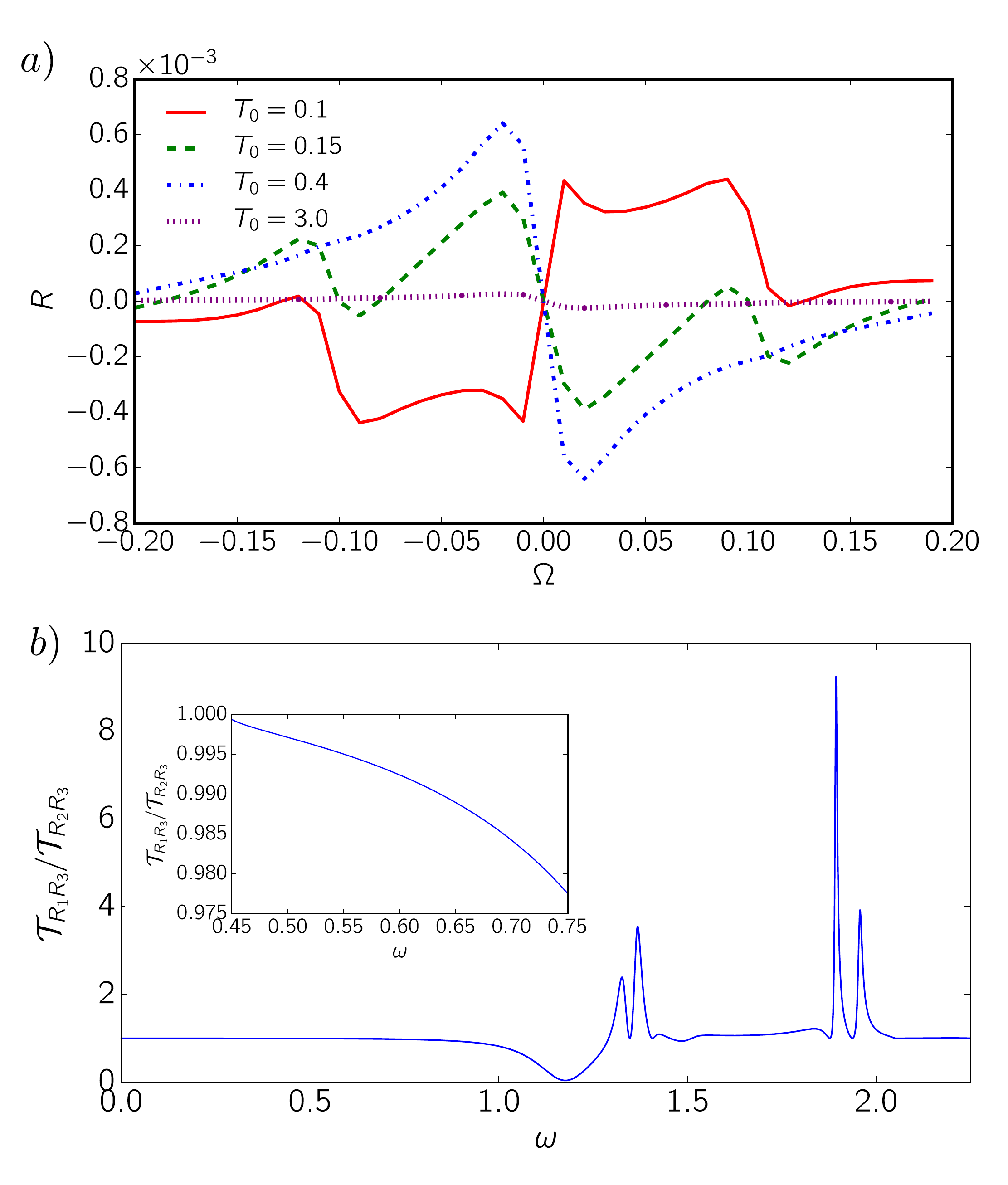}
\caption{(Color online) a) Plots of rectification $R$ versus angular velocity $\Omega$ for different average temperatures $T_0$. b) A plot of the transmission 
coefficient ratio $\frac{{\cal T}_{R_{1}R_{3}}}{{\cal T}_{R_{2}R_{3}}}$ versus frequency $\omega$, with the fixed angular velocity $\Omega = 0.03$. The inset 
shows this ratio from $\omega =$ 0.45 to 0.75. Other parameters are $N_b = 4$, $k_C = k = 1$ and $\Delta T=2.5\%T_0$.
\label{fig2}}
\end{figure}

\begin{figure}
\includegraphics[width=1\linewidth, angle=0]{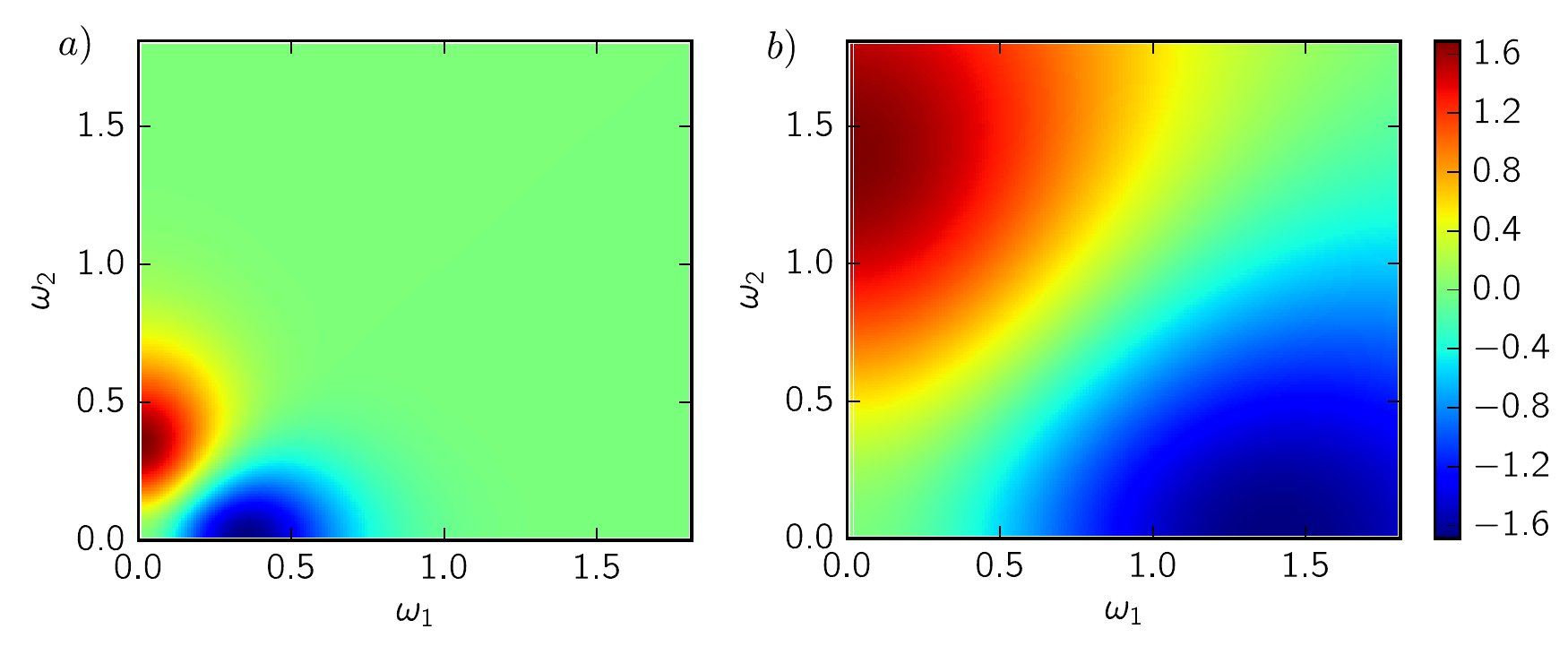}
\caption{(Color online) A density plot of the weight function $W(\omega_1,\omega_2,T_0)$ versus $\omega_1$ and $\omega_2$ for two different average 
temperatures a) $T_0=0.1$ and  b)$T_0=0.4$. Clearly, when the average temperature  $T_0$ increases, the dominant part of the magnitude of the weight-
function $|W(\omega_1,\omega_2,T_0)|$ moves to the higher frequency.  
\label{fig3}}
\end{figure}

Our numerical results for the rectification parameter $R$ versus the angular velocity $\Omega$ and different average temperatures $T_0$ are shown in 
Fig.~\ref{fig2}a. These data clearly demonstrate that the rectification effect is present once $\Omega\neq$0. Figure~\ref{fig2}a confirms that for high average 
temperatures $T_0$, corresponding to the classical limit, the rectification effect is suppressed (compare e.g. the curves for $T_0=0.1$ and $T_0=3$) as we have 
predicted in subsection~\ref{ssec:rectif}.

From  Fig. \ref{fig2}a we further see that the rectification parameter $R$ for a fixed angular velocity may change sign when the average temperature 
$T_{0}$ changes from $0.1$ to $0.4$. This can be understood from the perturbation result in Eq.~\eqref{eq:delta_I}. Figure~\ref{fig3} demonstrates 
that, when the average temperature $T_{0}$ changes from $0.1$ to $0.4$, the frequency region for which the absolute value of the weight function 
$|W\left(\omega_{1},\omega_{2},T_{0}\right)|$ is large will shifts towards high frequencies. Bearing this fact in mind, we are now able to explain 
the change of sign in the rectification. For example, for $\Omega = 0.03$, we find that the transmission ratio $\frac{{\cal T}_{R_{1}R_{3}}}
{{\cal T}_{R_{2}R_{3}}}$ initially decreases as a function of frequency while for larger frequencies it increases, see Fig.~\ref{fig2} b). As a result 
K (see Eq.~\eqref{kparameter}) changes sign from negative to positive so that the rectification parameter $R$ Eq.~\eqref{eq:delta_I} changes 
also sign from positive to negative.

Furthermore, we see that, the rectification direction is controlled by the sign of the angular velocity $\Omega$, thus allowing a greater flexibility 
to reconfigure ``on the fly" the direction of the heat current. This can be understood from the equality $\frac{{\cal T}_{R_{1}R_{3}}}{{\cal 
T}_{R_{2}R_{3}}}[\omega, \Omega]=1/(\frac{{\cal T}_{R_{1}R_{3}}}{{\cal T}_{R_{2}R_{3}}}[\omega, -\Omega])$ which is directly derived from 
Eq.~\eqref{Tsym} and Eq.~\eqref{Trelations}. Specifically, due to this equality, in the frequency region where $|W(\omega_1,\omega_2,T_0)|$ 
is large (for a fixed average temperature $T_0$), the behaviour of $\frac{{\cal T}_{R_{1}R_{3}}}{{\cal T}_{R_{2}R_{3}}}$ as a function of $\omega$ 
will turn from decreasing (increasing) function to increasing (decreasing) function when $\Omega\rightarrow -\Omega$. Thus the function $K$ 
(defined in Eq.~\eqref{kparameter}) and correspondingly the rectification parameter $R$ will change sign when we reverse the angular velocity 
$\Omega$.

\begin{figure}
\includegraphics[width=1\linewidth, angle=0]{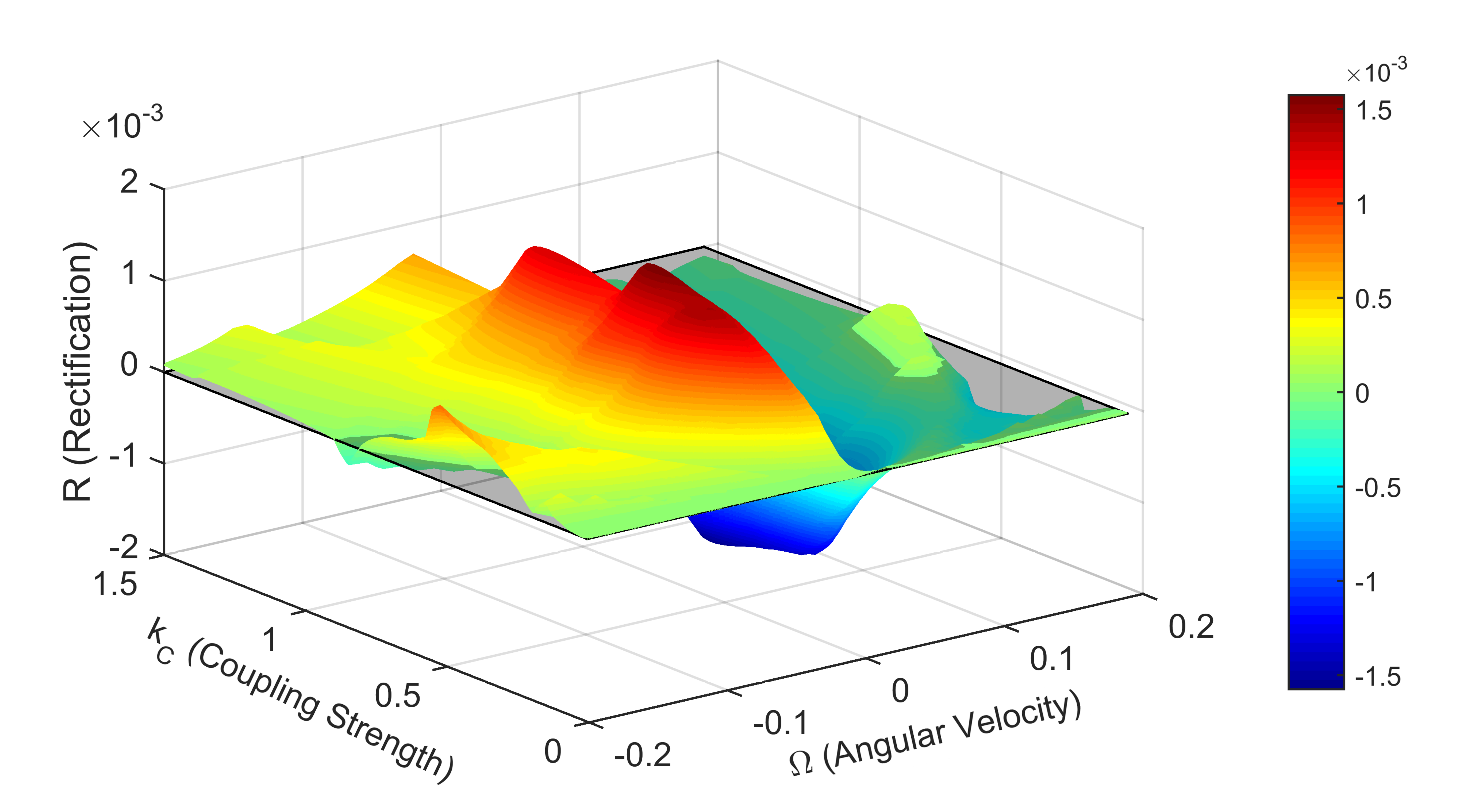}
\caption{(Color online) A 3D plot of rectification $R$ versus angular velocity $\Omega$ and spring constant $k_C$ in the center lattice. Other parameter are 
$N_b = 4$, $T_0=0.4$, $k = 1$ and $\Delta T=2.5\%T_0$. At $k_c\approx 0.65$ and $\Omega\approx \pm 0.02$ the rectification gets its maximum value. 
\label{fig4}}
\end{figure}

Finally, a 3D plot for the rectification $R$ versus $k_{C}$ and the angular velocity $\Omega$ is shown in Fig.~\ref{fig4}. In these calculations the spring constant
that couples the particles $R_1,R_2,R_3$ with the heat baths is taken to be $k=1$. We see that the maximum rectification of $R_{max}=1.6 \times 10^{-3}$ is observed for $k_C\approx 0.65$ and $\Omega\approx \pm 0.02$ near the origin. The non-monotonic behaviour of $R$ versus $k_c$ is associated with an impedance missmatch between the two chains
associated with the central juntion lattice (with coupling constant $k_c$) and the one-dimensional chain of the bath (which involves coupling constants $k=1$). 

\section{Conclusions}\label{sec:conclusions}

We have proposed the use of the Coriolis force in order to break time-reversal symmetry and induce thermal rectification in a ballistic three terminal nano-junction. Two of these terminals are attached to a hot and a cold reservoir, respectively, while the third one is attached to a probe reservoir whose temperature is self-consistently adjusted such that the net current towards this reservoir is zero. The nano-junction consists of a symmetric lattice which is placed on a rotating platform. Using non-equilibrium Green's 
function formalism we have calculated the rectification effect up to second-order in the temperature difference between the two reservoirs. We conclude that the 
Coriolis-induced rectification effect is of quantum mechanical nature and its direction and magnitude depend on $\Omega$. It will be interesting to extend this
study beyond the ballistic limit and investigate the effects of non-linearity and disorder. The latter phenomena can be important to any realistic structure where phonon-phonon interactions and structural imprefections are generally also present.

\end{document}